# New harms moderated by immersive experiences and breaks in them

EUGENE KUKSHINOV, University of Waterloo

This proposal highlights the potential real-world consequences of harmful experiences in immersive and embodied spaces due to presence, which moderates experiences, intensifying positive or negative content. While positive experiences enhance social interactions, negative content can lead to harm. Also, presence is not continuous and may break. Understanding the threshold between harmful experiences and breaks in presence is essential. This proposal underscores the need to explore whether the level of immersion or the nature of harm defines the boundary, emphasizing key discussion points for the workshop.

## 1 BACKGROUND

I am a media psychology researcher with a primary focus on the sense of presence, embodiment, and immersive experiences. One of the specific areas of my recent research is Social VR, which inherently enables various forms of embodied social interaction in virtual spaces. While this offers rich experiences, it also carries the potential for harm to users, depending on the nature and context of the simulation.

## 2 THE DISCUSSION PROPOSAL

As much as any other experience, harmful experiences in immersive and embodied spaces can translate to the real world because of presence, which is a perceptual illusion of non-simulation [3] and/or non-mediation [4]. This means that technology users to some extent do not recognize the role of the technology in their experiences. Presence is about the effect. It is the point of presence, embodiment, and/or immersive experiences, i.e., gaining real experiences and sensations from simulated or mediated events/activities. This is how VR training or VR exposure therapy works.

However, the nature of the experiences defines the direction of the effect. In other words, presence and/or embodiment are neither positive nor negative metrics, they are moderating variables. How users feel depends on the content. People would feel positive about being surrounded by friends or family, being in desired spaces, or doing something fun. If content is positive then it can be experienced as one. However, if it is not, then the experience can be harmful. Greater embodiment, presence, and immersion can lead to more intense experiences of harassment in Social VR ([1]) because presence moderates them, it does not create them by itself. Again, presence (in VR) is an illusion that something that is simulated feels like non-simulated. This may induce any traumatic or stressful content.

At the same time, presence is not a continuous state, it may (and always) break [2]. The reasons for breaks may differ, however, it is reasonable to assume that some disturbing content can break presence. Therefore, the question arises – where is the line between harmful experiences and breaks in presence? How much can an individual be immersed to still experience harmful content? Is it about the level of harm or the level of presence that defines it? This is the main area I would like to discuss during the workshop.